\documentstyle[12pt]{article} 
  \let\LARGE=\large
 \let\large=\normalsize

\begin{document}

\begin{titlepage}
\begin{flushright}
LPC-94-39\\
hep-ph/9411224\\
\end{flushright}
\begin{center} {\LARGE \bf
BOUNDS ON COMPOSITENESS FROM    \\
NEUTRINOLESS DOUBLE BETA DECAY \\}
\vspace*{0.8cm}
{Orlando PANELLA$^{1,2}$ and Yogendra N.~SRIVASTAVA$^{1,3}$}\\
\vspace*{1.2cm}
$^{1)}$~Dipartimento di Fisica dell' Universit\`a and
INFN Sezione di Perugia\\
Via A. Pascoli I-06100 Perugia, Italy.\\
\vspace*{1.0cm}
$^{2)}$~Coll\`ege de France, Laboratoire de
Physique Corpusculaire \\
11, Place Marcelin Berthelot, F-75231 Paris, Cedex 05, France.\\
\vspace*{1.0cm}
$^{3)}$~Physics Department, Northeastern University \\
Boston, Massachusetts 02115  \\
\vspace*{1.8cm}
{\bf   Abstract  \\ }
\end{center} \indent

Assuming the existence of a  heavy Majorana neutral particle
arising from a composite model scenario we
discuss the constraints imposed by present
experimental limits of  half-life neutrinoless double beta decay
$(0\nu\beta\beta ) $ measurements
on the coupling of the heavy composite neutrinos to the gauge bosons.
For neutrino masses $M_N = 1 $ TeV we obtain a rather weak
lower bound on the
compositeness scale:  $\Lambda \ge 0.23 $ TeV.

\end{titlepage}

\def\VACEV#1{\Big\langle 0 \Big\vert #1 \Big\vert 0 \Big\rangle}
\def\fqv#1{f_{q/V_#1}}
\def\goes#1{\buildrel  {#1}   \over \longrightarrow }

Heavy neutral Majorana particles  with masses in the
TeV region are predicted
in various theoretical models,  such as
superstring-inspired E$_6$  grand unification \cite{e6} or
left-right symmetric models \cite{lrs}.
In addition the possibility of a
fourth generation with a heavy neutral lepton, that could be
of Majorana
type, is not yet ruled out \cite{hill,datta}.

In this paper we discuss the possibility that  a heavy Majorana
neutrino might  arise from a composite model of  the ordinary
fermions \cite{barbieri}. Composite models, which describe
quarks and leptons as bound states of still more
fundamental particles, generally called preons, have been developed
as alternatives to overcome some of the theoretical problems
of the standard model
 \cite{hara}.
\par Although no completely consistent dynamical composite theory
has been found to date, various models have been proposed, and
one common, (inevitable), prediction of these models
is the existence of excited states of the known
quarks and leptons, much in the same way as the hydrogen atom
has a series of higher energy levels above the ground state.
The masses of the excited particles should not be much lower than the
compositeness scale ${\Lambda}$, which is expected to be at
least of the order of a TeV  according to experimental
constraints. For example  the search for
four-fermion contact interactions gives $\Lambda (eell) >
0.9-4.7 $ TeV  depending on the chirality of the coupling and
on the lepton flavour \cite{aleph,pdg}.
We expect therefore the heavy fermion masses to be
of the order of a few hundred GeV. The CDF experiment has
excluded excited quarks in the mass range $ 90-570 $ GeV
from $\gamma\, + $
jet and W + jet final states \cite{cdf}.
\par
Phenomenological implications of heavy fermions have been discussed
in the literature  \cite{cab,baur}
using weak isospin ($I_W$) and hypercharge ($Y$) conservation.
Assuming that such states are grouped in $SU(2) \times U(1)$
multiplets, since  light  fermions have $I_W=0,1/2$
and electroweak  gauge bosons have $I_W=0,1$,
to lowest order in perturbation theory, only
multiplets with $I_W \leq 3/2$ can be excited.
Also, since none of the gauge fields carry hypercharge, a
given excited multiplet can couple only to a light multiplet with
the same $Y$. In addition, current conservation forces  the coupling
of the heavy fermions to gauge bosons to be of the
magnetic moment type.
\par
We will only consider here the excited multiplet with
$I_W=1/2 \quad Y=-1$
\begin{equation}
{\cal  E} = {N \choose E}
\end{equation}
which can couple to the light  left multiplet
\begin{equation}
\ell_L = {\nu_L \choose e_L} ={{1-\gamma_5} \over 2}
{\nu \choose e}
\end{equation}
through the gauge fields ${\vec W}^{\mu}\hbox  {and }  B^{\mu}$,
with the additional assumption that N is a neutral Majorana
fermion.
\par
\par
In terms of the physical gauge fields $
W^{\pm}_{\mu}=(1 / \sqrt 2) \Bigl( W^1_{\mu} \mp \, i \,
W^2_{\mu} \Bigr)
$
the relevant effective interaction can be expressed as
\begin{eqnarray}
{\cal L}_{eff}& = & \bigl({ g f \over {\sqrt 2} \Lambda} \bigr)
\Bigl\{
\Bigl( {\overline N}
\sigma^{\mu \nu} {1-\gamma_5 \over 2} \, e \Bigr)
\, \partial_{\nu} W_{\mu}^+\nonumber \\  & \,
& \qquad \qquad +\Bigl( {\overline E} \sigma^{\mu \nu} {1-\gamma_5
\over 2}
\, \nu \Bigr) \, \partial_{\nu} W^-_{\mu} + h.c. \Bigl\}
\, + \,  \hbox{neutral currents}.
\end{eqnarray}
where $ f $ is a dimensionless coupling constant,
$\Lambda$ is the compositeness scale,
and ${\vec \tau}$ are the Pauli $SU(2)$
matrices, and the rest of the notation is as usual in the standard
model.
An extension to quarks and other multiplets, with
a detailed discussion of the spectroscopy of the excited particles
can be found in Ref. \cite{pan}.

Regarding the experimental mass limits on the heavy Majorana
neutrinos
from pair production, $Z \to N \bar{N}$, we have
$ M_N > 34.6 $ GeV at 95\,\% c.l.,
which  has been deduced from
the Z line shape measurements \cite{aleph2}, and which
is independent of the
decay modes.
More stringent limits $\approx 90 $ GeV come  from single
excited neutrino
production, $Z \to N \nu $,
through the transition
magnetic coupling, but these do depend on assumptions regarding
the branching
ratio of the decay channel chosen \cite{pdg,hera,aleph2}.

In practical calculations of production cross sections and decay rates
of excited states, it has been customary \cite{chia,hagi,baur}
to assume that
the dimensionless coupling $f$ in Eq. (3) is of order unity. However
if we assume that the excited
neutrino is of Majorana type, we have to verify that this
choice is compatible with present experimental limits on neutrinoless
double beta decay
($ 0\nu \beta \beta$):

\begin{equation}
A(Z) \to A(Z+2) + e^- + e^-
\end{equation}
a nuclear decay, see Fig. 1, that has attracted
much attention both from particle and nuclear physicists
because
of its potential to expose lepton number violation.
More generally,  it is expected to give interesting insights about
certain gauge theory parameters such as leptonic charged mixing matrix,
neutrino masses etc.
The process in Eq.(4), which can only proceed via the exchange of
a massive Majorana neutrino, has been experimentally searched
for in a number
of nuclear systems \cite{exp}
and has also been extensively studied from the
theoretical side \cite{hax,verga,klapdor}.

We will consider here  the decay
\begin{equation}
^{76}\hbox{Ge} \to \, {^{76}\hbox{Se}} + 2e^-
\end{equation}
for which we have from the Heidelberg-Moscow $\beta\beta$-experiment
the recent limit  \cite{balysh} ($T_{1/2}$
is the half life =
log2 $ \times $ lifetime)
\begin{equation}
T_{1/2} \,  (\, ^{76}\hbox{Ge} \to \ {^{76}\hbox{Se}} + 2e^-)
\geq 1.95 \times 10^{24}\, \hbox{yr} \quad 90\,\% \hbox{ c.l.}
\end{equation}

In the following  we estimate the constraint imposed by the above
measurement
on the coupling $(f/\Lambda)$ of the heavy composite neutrino, as
given by Eq. (3).
The fact that
neutrinoless double beta decay measurements
might constraint composite
models, was also discussed in ref.\cite{barbieri} but
within the framework of a particular model and referring
to a heavy Majorana neutrino with the usual $\gamma_\mu$ coupling.

The transition amplitude of $0\nu \beta \beta$ decay is calculated
according to the interaction lagrangian:
\begin{eqnarray}
{\cal L}_{int}& =& {g \over 2 \sqrt{2}}\left\{
{f \over \Lambda}
\bar{\psi}_e(x) \sigma_{\mu\nu}(1+\gamma_5)\psi_N(x)
\partial^\mu W^{\nu (-)}(x)\right.  \nonumber \\
& \qquad & \qquad \qquad  \left. +\cos\theta_C J^h_\mu (x)
W^{\mu (-)}(x) + h.c. \right\}
\end{eqnarray}
where $\theta_C$ is the Cabibbo angle ($\cos\theta_C = 0.974$ ) and
$J_\mu^h$ is the hadronic weak charged current
\begin{eqnarray}
J_\mu^h (x) & = &\sum_k j_\mu (k) \delta^3 ({\bf x} - {\bf r}_k )
\nonumber \\
j_\mu(k) & = & {\overline {\cal N}}({\bf r}_k) \gamma_\mu (
f_V -f_A\gamma_5)\tau_-(k){\cal N}({\bf r}_k)
\end{eqnarray}
and where ${\bf r}_k $ is the coordinate of the k-th nucleon,
$ {\cal N} = \left (
{\psi_p \atop \psi_n } \right ) $ and $\tau_-(k) = (1/2)
(\tau_1(k)-i\tau_2(k) )$
is the step down operator for the isotopic spin, (${\vec\tau}(k)$
is the
matrice describing the isotopic spin of the k-th nucleon).
We emphasize that in Eq. (7)
we have a $\sigma_{\mu\nu} $ type
of coupling as opposed to the $\gamma_\mu $ coupling
so far encountered in all $0\nu\beta\beta $ decay  calculations.

For simplicity, we carry out our analysis
assuming that there are no additional contributions to $0\nu\beta\beta$
decay from light Majorana neutrinos, right handed currents
or other heavy Majorana neutrinos originating from another source.

The transition amplitude is then
\begin{eqnarray}
S_{fi}& =& (cos\theta_C)^2 ({g \over 2 \sqrt{2}})^4
\left( {f \over \Lambda}\right)^2 ({1\over 2})
\int\, {d^4k \over (2\pi)^4}\, d^4x\, d^4y e^{-ik\cdot (x-y)}
\times \nonumber \\
&\, &{1\over \sqrt{2}}(1-P_{12})
{\bar u}(p_1)\sigma_{\mu\lambda}(1+\gamma_5)
{\not\! k +M_N \over k^2 -M_N^2} (1+\gamma_5)
\sigma_{\nu\rho}v(p_2) \times
\nonumber \\
&\ & \bigl[ F(Z+2,\epsilon_1)
F(Z+2,\epsilon_2)\bigr]^{1/2} e^{ip_1\cdot x}
e^{ip_2 \cdot y} f_A((k-p_1)^2) f_A((k+p_2)^2) \times\nonumber \\
& \ &  (k-p_1)^\lambda(k+p_2)^\rho
{<f|J^\mu_h(x) \ J^\nu_h(y)|i>
\over [(k-p_1)^2 -M_W^2] [(k+p_2)^2-M_W^2]}
\end{eqnarray}
where $(1-P_{12})/\sqrt{2}$ is the antisymmetrization operator due to
the production of two identical fermions,
the functions $F(Z,\epsilon$)
are the well known Fermi functions \cite{rospri}
that describe the distorsion of
the electron's plane wave due to the nuclear Coulomb field
($\epsilon_i$ are the electron's kinetic
energies in units of $m_ec^2$),
\begin{eqnarray}
F(Z,\epsilon)&=&\chi(Z,\epsilon)
\frac{\epsilon +1}{[\epsilon(\epsilon +2)]^{1/2}} \\
\chi(Z,\epsilon)&\approx & \chi^{R.P.}(Z)=
\frac{2\pi \alpha Z}{1-e^{-2\pi\alpha Z}}\quad \hbox{(Rosen-Primakoff
aproximation)} \nonumber
\end{eqnarray}
and the nucleon form factor,
\begin{equation}
f_A(q^2)={1\over (1+{|{\bf q}|^2/ m_A^2})^2}
\end{equation}
with $m_A=0.85 $ GeV,
is introduced to take into account the finite size of the nucleon,
which is known to give important effects for the heavy neutrino case.

As is standard in such calculations,
we make the following approximations:
  \cite{hax,verga}:\\
$i$) the hadronic matrix element is evaluated within the closure
approximation
\begin{equation}
<f|J^\mu_h(x) \ J^\nu_h(y)|i>\,  \approx e^{i(E_f-<E_n>)x_0}
 e^{i(<E_n> - E_i)y_0} <f|J^\mu_h({\bf x}) \ J^\nu_h({\bf y})|i>
\end{equation}
where $<E_n>$ is an average excitation
energy of the intermediate states.
This allows one to perform the integrations over $k_0, x_0, y_0$
in Eq. (9);\\
$ii$) neglect the external momenta $p_1$, $p_2$ in the propagators
and use the long wavelength approximation $e^{-i{\bf p}_1\cdot {\bf x}}
= e^{-i{\bf p}_2 \cdot {\bf y}} \approx 1 $; \\
$iii$) the average virtual neutrino momentum
$<|{\bf k}|> \,  \approx \, 1/R_0 = 40$
MeV is much  larger than the typical
low-lying excitation energies, so that, $k_0 = E_f +E_1 -<E_n> $
can be neglected relative to {\bf k};\\
$iv$) the effect of W and N propagators can be neglected since
$M_W \approx 80 $\ GeV is much greater
than $|{\bf k}|$ in the region where
the integrand is large, and we are interested in heavy neutrino masses
$M_N \gg M_W $.

Using the same notation as in Ref.  \cite{verga} we arrive at
\begin{eqnarray}
S_{fi} &=& (G_F\cos\theta_C)^2{f^2 \over \Lambda^2}  {1\over 2}
2\pi \delta(E_0 - E_1 - E_2) \times \nonumber \\
& \ & {1\over \sqrt{2}}(1-P_{12}){\bar u}(p_1)\sigma_{\mu i}
\sigma_{\nu j} (1+\gamma_5)v(p_2)
\bigl[ F(Z+2,\epsilon_1)F(Z+2,\epsilon_2)\bigr]^{1/2} \times
\nonumber \\
& \ & M_N \sum_{kl} I_{ij} <f|j^\mu(k)j^\nu(l)|i>
\end{eqnarray}
where $I_{ij}$ is an integral over the virtual neutrino momentum,
(${\bf r}_{kl}=  {\bf r}_k -{\bf r}_l\, ,
r_{kl}= |{\bf r}_k -{\bf r}_l|\, ,
x_{kl}=m_Ar_{kl}) $
\begin{eqnarray}
I_{ij}& = & \frac {1} {M_N^2} \int\, \frac {d^3 {\bf k} } {(2\pi)^3} \,
e^{i{\bf k}\cdot {\bf r}_{kl}}\, \frac {(-k_i k_j)}
{(1+|{\bf k}|^2/m_A^2)^4}\,
\nonumber \\
& = & \frac {1}{4\pi} \frac {m_A^4}{M_N^2}\frac{1}{r_{kl}} \left\{
-\delta_{ij} F_A(x_{kl}) +\frac {({\bf r}_k)_i({\bf r}_l)_j}
{r_{kl}^2} F_B(x_{kl})
\right\} \\
\hbox{with:}&\,  & \,  \nonumber \\
& \, & F_A(x)= \frac {1}{48} e^{-x} \, (x^2 +x)    \nonumber \\
& \, & F_B(x)= \frac {1}{48} e^{-x} \, x^3
\end{eqnarray}

Since $I_{ij}$ is a symmetric tensor, we can make the replacement
$\sigma_{\mu i}\sigma_{\nu j} \to (1/2)\{\sigma_{\mu i},\sigma_{\nu j}\}
= \eta_{\mu \nu} \eta_{ij} -\eta_{i\nu} \eta_{i\mu} +
i \gamma_5 \epsilon_{\mu i\nu j} $. Then, using the nonrelativistic
limit of the nuclear current
\begin{equation}
j_\mu(k) = \left\{
\begin{tabular}{ll}
$f_V\tau_-(k)$ & if $\mu=0$\\
$-f_A\tau_-(k) (\sigma_k)_i$ & if $\mu=i$
\end{tabular}
\right.
\end{equation}
($\vec{\sigma}_k $ is the spin matrice of the k-th nucleon)
we arrive, with straightforward algebra, at
\begin{eqnarray}
S_{fi} & = & M_{fi} \, 2\pi \, \delta (E_0 -E_1 -E_2)\nonumber \\
M_{fi} & = &(G_F\cos\theta_C)^2\frac {1}{4}
\frac{-1}{2 \pi} \frac{f_A^2}
{r_0 A^{1/3}} \, l \, <m>
\end{eqnarray}
where we have defined
\begin{eqnarray}
l & =& {1\over \sqrt{2}}(1-P_{12}){\bar u}(p_1)
(1+\gamma_5)v(p_2)
\bigl[ F(Z+2,\epsilon_1)F(Z+2,\epsilon_2)\bigr]^{1/2}\nonumber \\
<m>& = & m_e \eta_N <f\,|\,\Omega\,|\,i> \nonumber \\
\eta_N & = & \frac{m_p}{M_N} m_A^2 \left ( \frac{f}{\Lambda} \right )^2
\nonumber \\
\Omega & = & \frac {m_A^2}{m_p m_e} \sum_{k\neq l} \tau_-(k)\tau_-(l)
\frac{R_0}{r_{kl}} \left [ \left (\frac{f_V^2}{f_A^2}-{\vec \sigma}_k
\cdot {\vec \sigma}_l \right )
(F_B(x_{kl}) -3F_A(x_{kl}))\right. \nonumber \\
& \, &\hbox{\hspace{1.5cm}}\left.  -{\vec \sigma}_k
\cdot {\vec \sigma}_l\, F_A(x_{kl}) +
\frac {  {\vec \sigma}_k\cdot{\bf r}_{kl} \,
{\vec \sigma}_l\cdot{\bf r}_{kl} } {r_{kl}^2} F_B(x_{kl})
\right ]
\end{eqnarray}
and $R_0=r_0A^{1/3}$ is the nuclear radius ($r_0 =1.1 $ fm).

The new result here is the nuclear operator $\Omega $ which
is substantially different from those so far encountered in
$0\nu\beta\beta $
decays, due to the $ \sigma_{\mu\nu} $ coupling of the heavy neutrino
that we are considering.
The decay width is obtained upon integration over the density
of final states of the two-electron system
\begin{equation}
d\Gamma =  \sum_{final\, spins}
{|M_{fi}|^2}\,  2\pi\delta (E_0 -E_1 -E_2)
\frac {d^3{\bf p}_1} {(2\pi)^3 \, 2E_1}
\frac {d^3{\bf p}_2} {(2\pi)^3 \, 2E_2}
\end{equation}
and the total decay rate $\Gamma$ can be cast in the form
\begin{eqnarray}
\Gamma & = & (G_F \cos\theta_C)^4\frac{(f_A)^4 \,m_e^7
\,|\eta_N|^2} {(2\pi)^5
r_0^2 A^{2/3}} \, f_{0\nu}(\epsilon_0 , Z) \,
|\Omega_{fi}|^2 \\
f_{0\nu} & = & \xi_{0\nu} f_{0\nu}^{R.P.}\\
f_{0\nu}^{R.P.}& = & |\chi^{R.P.}(Z+2)|^2 \frac {\epsilon_0}{30}
(\epsilon_0^4+10\epsilon_0^3 +40\epsilon_0^2+60\epsilon_0 +30 )
\end{eqnarray}
where, $\Omega_{fi} =\, <f|\Omega |i> $, $\epsilon_0 $
is the kinetic energy of the two electrons in units
of $m_ec^2$, and $\xi_{0\nu}$ is a numerical factor that corrects
for the Rosen-Primakoff
approximation  \cite{verga} used in deriving the analytical
expression of $f_{0\nu}^{R.P.}$.
For the decay considered in Eq.(5), we have  \cite{verga}
 $ \xi_{0\nu} = 1.7 $ and $\epsilon_0 = 4. $
The half-life is finally written as
\begin{eqnarray}
T_{1/2}& = & \frac {K_{0\nu}\, A^{2/3}}
{f_{0\nu}\, |\eta_N|^2 \, |\Omega_{fi}|^2}\\
K_{0\nu} & = & (\log 2) \frac {(2\pi)^5} {(G_F\cos\theta_Cm_e^2)^4}
\frac{(m_e r_0)^2}{m_e f_A^4} = 1.24 \times \, 10^{16} \, \hbox{yr}
\nonumber
\end{eqnarray}

Combining Eq. (23) with the experimental limit given for the decay
considered in Eq. (5),
we obtain a constraint on the dimensionless coupling $f $
\begin{equation}
|f| \leq \left ( \frac {M_N \Lambda^2}{m_p m_A^2} \right )^{1/2}
\left [ \frac {K_{0\nu} \, A^{2/3}}{ 1.4 \times 10^{24} \,
\hbox{yr} \times f_{0\nu}(Z,\epsilon_0)}\right ]^{1/4}
\frac {1} {\, \, |\Omega_{fi}|^{1/2}}
\end{equation}

Given the heavy neutrino mass $M_N $
and the compositeness scale $\Lambda $, we only need to evaluate
the nuclear matrix element $\Omega_{fi} $ to know the
upper bound on the value of $|f|$ imposed by neutrinoless
double beta decay.

The evaluation of the nuclear matrix elements
was in the past regarded as the principal source of uncertainty
in $0\nu\beta\beta$ decay calculations, but
the recent high-statistics measurement \cite{2nu}
of the allowed $2\nu\beta\beta$
decay, a second order weak-interaction $\beta $ decay, has
shown that nuclear physics can provide a very good description of
these phenomena, giving high reliability to the constraints imposed
by $0\nu\beta\beta$ decay on non-standard model parameters.

Since we simply want
to  estimate  the order of magnitude of the constraint
in Eq. (24) we will evaluate the nuclear matrix element
only approximatively.
First of all the expression of the nuclear operator in Eq. (18)
is simplified making the following replacement
\begin{equation}
\frac { r_{kl}^i r_{kl}^j} {r_{kl}^2}\,  \to \,
< \frac { r_{kl}^i r_{kl}^j} {r_{kl}^2}\, > \, \to \,
\frac {1}{3} \delta_{ij}
\end{equation}
The operator $\Omega $ becomes then
\begin{equation}
\Omega  \approx  \frac {m_A^2}{m_p m_e}(m_A R_0)
\sum_{k\neq l} \tau_-(k)\tau_-(l)
\left (\frac{f_V^2}{f_A^2}-
\frac {2}{3} {\vec \sigma}_k
\cdot {\vec \sigma}_l \right )  F_N(x_{kl})
\end{equation}
where $
F_N = (1/x)(F_B -3F_A)=(1/48)e^{-x}(x^2-3x-3)$
with $F_B$ and $F_A$ given in Eq.(15).

Since we are interested in deriving the lowest possible
upper bound on $\vert f\vert $ given by Eq. (24), let us
find the maximum absolute value of the
nuclear matrix element of the
operator $\Omega $ in Eq.(18):
\begin{equation}
\vert \Omega_{fi} \vert \leq\frac{m_A^2}{m_p m_e}(m_A R_0)
\vert F_N(\bar{x})\vert \left\{\frac{f_V^2}{f_A^2}
 \vert M_F \vert +\frac{2}{3}\vert
M_{GT} \vert \right\}
\end{equation}
where $M_F=\ <f\vert \sum_{k\neq l}\tau_-(k)\tau_-(l) \vert i> $ and
$M_{GT} =\ <f\vert  \sum_{k \neq l}\tau_-(k)\tau_-(l)\vec \sigma_k \cdot
\vec \sigma_l\vert i>  $ are respectively the matrix elements of
the Fermi and Gamow-Teller
operators whose numerical values for the nuclear
system under consideration are  \cite{hax,verga},
$M_F = 0 $ and $ M_{GT} = -2.56$.
Inspection of the radial function
$F_N $ (for $x \geq 0 $) shows that its maximum absolute value
is  attained at $x = 0$.
In Eq. (27) we have evaluated $F_N$ at $x = 2.28\,
(r_{kl} = 0.5 $ fm).
This value of $r_{kl}$ corresponds to the typical internuclear distance
at which short range nuclear correlations become important  \cite{hax},
so that the region $x \leq 2.28 $ does not give contributions to the
matrix element of the nuclear operator.
We thus find
\begin{equation}
\vert \Omega_{fi} \vert \leq  0.6 \times 10^3
\end{equation}
which together with Eq. (24) gives the {\it conservative}
upper bound on $\vert f \vert $ shown in Fig. 2 as a function of $M_N$
for $\Lambda = 1 $ TeV.

In particular, we see that the choice $\vert f \vert \approx 1 $ is
compatible with bounds imposed by
experimental limits on neutrinoless
double beta decay rates.
We emphasize that our bound on $\vert f \vert $
is conservative, and an exact evaluation of the nuclear
matrix element will give an even higher lower bound.

We also note that Eq. (24) can alternatively
be used to give a lower bound on $\Lambda $ as a function of $M_N$
(assuming $\vert f \vert =1 $ ). This is shown in Fig. 3 where we can
see that the lower bound on the compositeness scale coming from
$0\nu\beta\beta$ decays is rather
weak: $\Lambda > 0.23 $ TeV at $M_N = 1$ TeV.
In table I we summarize our bounds for some values of the excited
Majorana neutrino mass.
We remark that, as opposed to the case of  bounds
coming from the direct search
of excited particles, our constraints on
$\Lambda $ and $\vert f  \vert $
{\it do not depend } on  any assumptions regarding the branching ratios
for the decays of the heavy particle.

To obtain more stringent bounds, we need  to improve
on the measurements of $0\nu\beta\beta$ half-life.
However, our bounds c.f. Eq. (24) on ($\vert f \vert $ or $\Lambda$ )
depend on the experimental $T_{1/2}$ lower limit  only
weakly ($\propto T_{1/2}^{\pm 1/4}$) so that to
obtain an order of magnitude more stringent bound we need
to push higher, by a factor of $10^4$, the lower bound
on $T_{1/2}$.

We should bear in mind, however, that
the simple observation of a few $0\nu\beta\beta$ decay events,
while unmistakably proving lepton number violation and the existence of
Majorana neutrals,
will not be enough to  uncover the originating mechanism (including the
one discussed here). In order to disentangle the various models,
single electron spectra will be needed, which would require high
statistics experiments and additional theoretical work.

\newpage
\begin{center}
{\bf ACKNOWLEDGMENTS}
\end{center}
\indent
This work was partially supported by the U.S. Department of Energy and
the Italian Institute for Nuclear Physics (Perugia).
One of us (O.P.) wishes to thank the Italo-Swiss foundation
``Angelo della Riccia''  and the University of Perugia (Italy)
for financial support.
He also would like to thank
C. Carimalo for useful discussions and
the Laboratoire de Physique Corpusculaire,
Coll\`ege de France, Paris,
where
this work was partially completed, for the very kind hospitality.
\newpage
\begin{center}
{\bf TABLE CAPTIONS}
\end{center}
\begin{itemize}
\item[{\bf [Table I]}]
Lower bounds on $\Lambda$ with $\vert f \vert $ = 1,
and upper bounds on $\vert f \vert $ with
$\Lambda = 1$ TeV, for different
values of the heavy neutrino mass $M_N$.
\end{itemize}
\vspace{2.0cm}
\begin{center}
{\bf FIGURE CAPTIONS}
\end{center}

\begin{itemize}
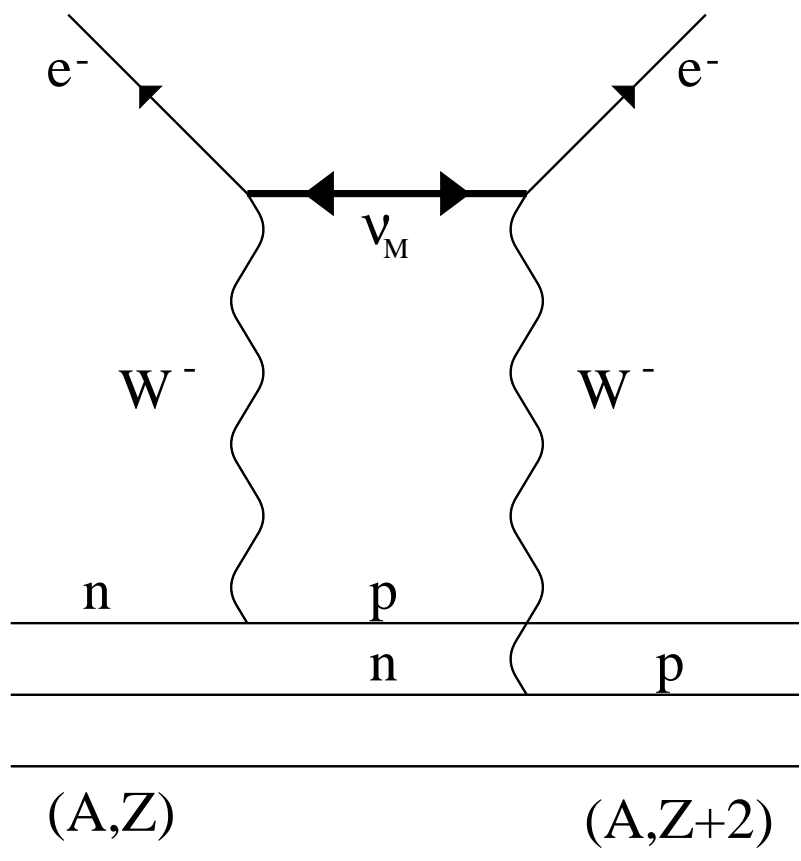
\item[{\bf [Fig. 1]}]
Schematic illustration of neutrinoless double beta decay $0\nu \beta
\beta $ via the exchange of a Majorana neutrino.
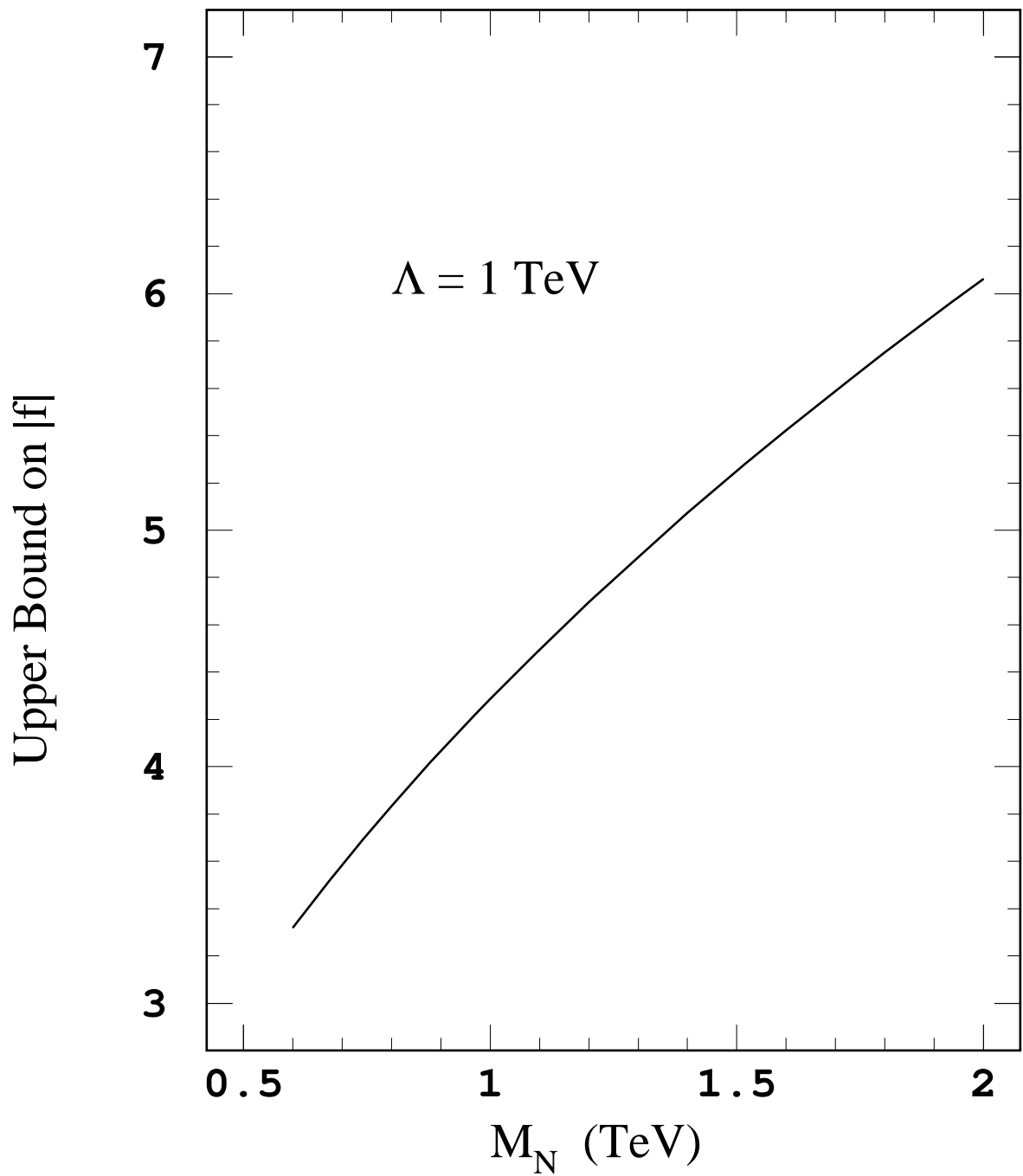
\item[{\bf [Fig. 2]}]
Conservative upper bound on $\vert f \vert $ versus the heavy Majorana
neutrino mass $M_N$, at $\Lambda = 1$ TeV.
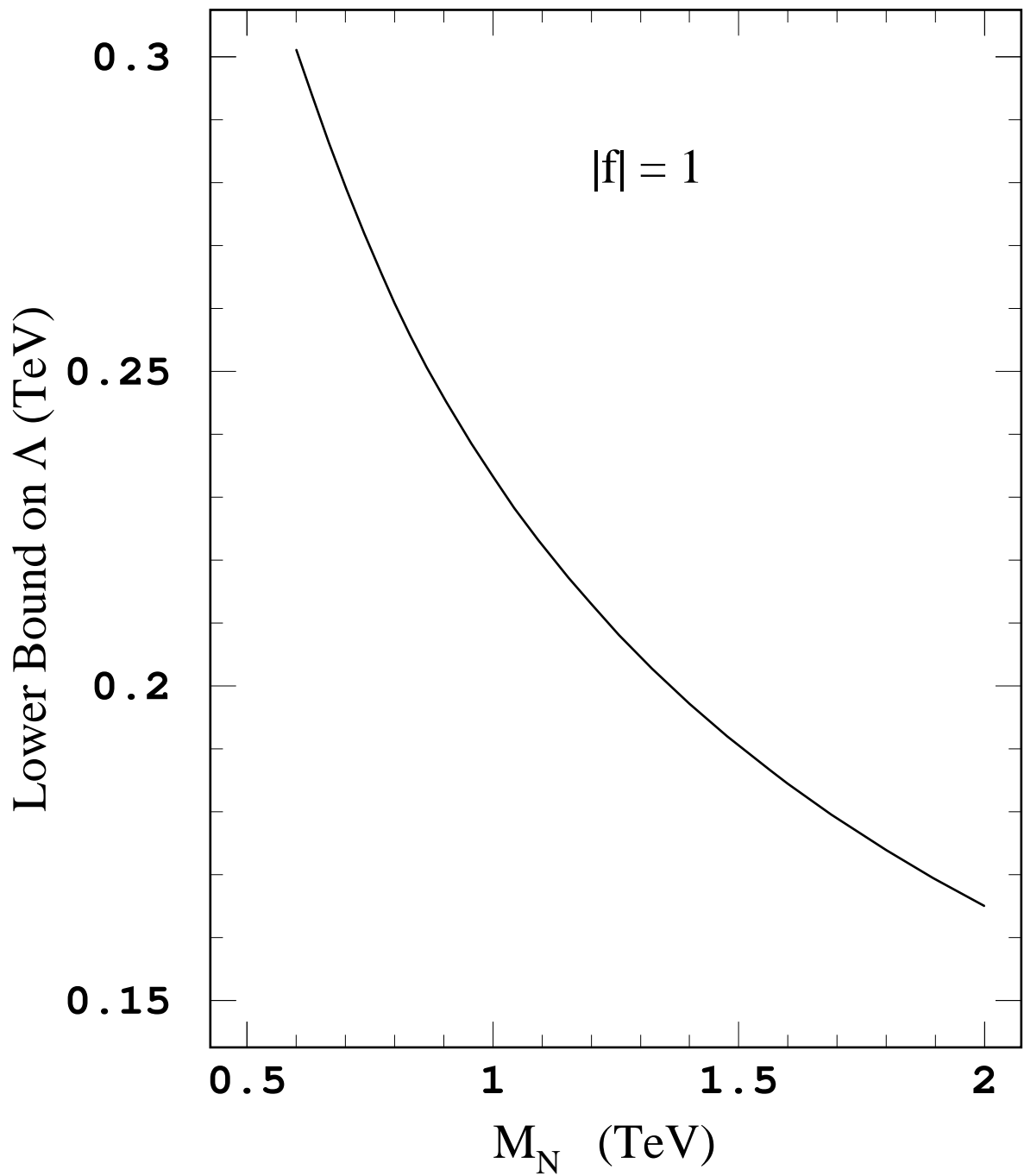
\item[{\bf [Fig. 3]}]
Conservative lower bound on $\Lambda $ versus
the heavy Majorana neutrino
mass $M_N$ for $\vert f \vert  = 1 $.
\end{itemize}

\newpage

\vspace{2.0cm}
\begin{center}

{\bf TABLE I}

\vspace{1.cm}
\begin{tabular}{|c|c|c|c|c|c|c|c|c|c|}
\hline
$M_N \hbox{(TeV)}$ & & 0.6 & 0.8 & 1.0 & 1.2 & 1.4 & 1.6 & 1.8 & 2.0 \\
\hline
$\Lambda\, \hbox{(TeV)} > $ &$[\vert f \vert = 1] $
& 0.30 & 0.26 & 0.23 & 0.21 & 0.20 & 0.18
& 0.17 & 0.16 \\
\hline
$\vert f \vert < $ &$[\Lambda = 1 \, \hbox{TeV}] $
& 3.3 & 3.8 & 4.3 & 4.7 & 5.1 & 5.4 & 5.7 & 6.1 \\
\hline
\end{tabular}
\end{center}

\end{document}